\documentclass[10pt,twocolumn,preprintnumbers,amsmath,amssymb,nofootinbib
,superscriptaddress]{revtex4}

\def\VEV#1{\left\langle #1\right\rangle}

\def\beq{\begin{equation}}  
\def\eeq{\end{equation}}  
\def\bea{\begin{eqnarray}}  
\def\eea{\end{eqnarray}}  
\def\half{\frac{1}{2}}  
  
\def\bq{\begin{quote}}  
\def\eq{\end{quote}}

\def\half{\frac{1}{2}}       

\relax  

\newskip\humongous \humongous=0pt plus 1000pt minus 1000pt

\newif\ifdtup

\def\oldreffmt#1{\rlap{[#1]} \hbox to 2\parindent{}}

\def\figfmt#1{\rlap{Figure {#1}} \hbox to 1in{}}  
  
\def\beq{\begin{equation}}  
\def\eeq{\end{equation}}  
\def\bea{\begin{eqnarray}}  
\def\eea{\end{eqnarray}}  
\def\half{\frac{1}{2}}  
  
\def\bq{\begin{quote}}  
\def\eq{\end{quote}}

\def\half{\frac{1}{2}}       

\relax  

\newdimen\tdim  
\tdim=\unitlength  
\def\bar{\overline}

\newcommand{\be}{\begin{equation}}  
\newcommand{\ee}{\end{equation}}  
\newcommand{\ba}{\begin{array}}  
\newcommand{\ea}{\end{array}}

\begin{document}
\preprint{FERMILAB-PUB--18--008T}
\preprint{OUTP-17-17P}

\title {Inertial Spontaneous Symmetry Breaking\\and Quantum Scale Invariance
}

\author{Pedro G. Ferreira}
\email{pedro.ferreira@physics.ox.ac.uk}
\affiliation{Astrophysics, Department of Physics\\
University of Oxford, Keble Road\\
Oxford OX1 3RH\\$ $}
\author{Christopher T. Hill}
\email{hill@fnal.gov}
\affiliation{Fermi National Accelerator Laboratory\\
P.O. Box 500, Batavia, Illinois 60510, USA\\$ $}
\author{Graham G. Ross}
\email{g.ross1@physics.ox.ac.uk}
\affiliation{Rudolf Peierls Centre for Theoretical Physics, \\
University of Oxford, 1 Keble Road\\
Oxford OX1 3NP\\$ $}

\date{\today}

\begin{abstract}

Weyl invariant theories of scalars and gravity 
can generate all mass scales spontaneously, 
initiated by a dynamical process of ``inertial spontaneous symmetry breaking" that does not involve a potential.
This is dictated by the structure of the Weyl current, $K_\mu$, and
a cosmological phase during which the universe expands and the Einstein-Hilbert 
effective action is formed.  Maintaining exact Weyl
invariance in the renormalized quantum theory is straightforward 
when renormalization conditions are
referred back to the VEV's of fields in the action of the theory, which implies 
a conserved Weyl current. We do not require scale invariant regulators.
We illustrate the computation
of a Weyl invariant Coleman-Weinberg potential.  
\end{abstract}

\maketitle

\maketitle
\section{Introduction}

The discovery of the Higgs boson with the appearance of a fundamental, 
point-like, scalar field, unaccompanied by a natural custodial
symmetry, has led many authors in search of new organising principles to turn 
to scale symmetry. In particular,
Weyl symmetry \cite{Weyl} in conjunction with gravity may provide a modern context 
for fundamental scalar fields
and a foundational symmetry for physics \cite{shapo}\cite{GB}\cite{authors}\cite{Englert:1976ep}\cite{S3}\cite{FHR}.
Scale or Weyl symmetry, like many of the flavour symmetries seen in nature, 
must be broken. Often this breaking is treated spontaneously, implemented for 
scale invariant potentials via the Coleman Weinberg (CW) mechanism of dimensional
transmutation \cite{Coleman:1973jx}. 

In this paper we focus on the well known Weyl current
which has been studied by many of the previous authors 
listed above (e.g. see \cite{shapo}). However, we
emphasise that, underlying these ideas there is 
a new way to break scale symmetry that {\it does not employ a potential}. While this 
mechanism is implicit in the many approaches taken to spontaneously generating the 
Planck scale, it seems not to have been codified prior to ref \cite{FHR}. This 
mechanism is a direct consequence of the structure of the Weyl scale current.  We call 
this {\em inertial spontaneous scale symmetry breaking.}

By inertial spontaneous scale symmetry breaking,  we presently follow 
the condensed matter parlance, where 
spontaneous symmetry breaking (SSB) represents the difference between an ``ordered'' state 
from an ``disordered'' one.  Initially, we expect local fluctuations in fields
that break scale, e.g., nonzero $\phi_i(x)$ or $\partial\phi_i(x)$, etc.
These are analogous to local magnetic spins in a macroscopic spin system at high temperature
and they do break scale symmetry locally, but they do not represent an ordered state. 
For that we need an ``order parameter'' that evolves
to become macroscopically constant in space and time,
like the constant magnetization in the spin system as it cools to the groundstate.
The order parameter must capture any and all symmetry breaking. 

In the present paper, working in a ``Jordan frame'' we identify the conserved Weyl current $K_\mu$.
We find that in any Weyl invariant theory this current is always a derivative
of a scalar quantity,  $K_\mu=\partial_\mu K$ where $K$ is the ``kernel.''\footnote{This is a theorem, and it isn't
hard
to see that it 
holds in $R^2$ and Weyl gravity generalizations, but the proof is beyond the scope
of our present discussion.}
Owing to this structure of the conserved Weyl current, $D_\mu K^\mu = 0$, we 
are guaranteed that the system 
will dynamically evolve in an expanding universe such that scale charge density will
evolve to zero, $K_0\rightarrow 0$.   This is just covariance, like the dilution of
a conserved electric charge density or a magnetic field during general expansion.

It then follows
that the kernel, $K\rightarrow \bar{K}$, is constant.  Essentially all short
distance 
initial scale fluctuations are stretched out to become
a constant value of $K=\bar{K}$, and the scale symmetry is broken. 
$K$ is the order parameter of the SSB since 
it intimately connects with the dynamics.
At late times we have, for $N$-fields, $K(\{\phi_i\})=\bar{K}(\{\phi_i\})\exp(\sigma/f)$
. Here $\sigma$ is the dilaton  and  $\bar{K}(\{\phi_i\})$ is a constraint that
reduces the $N$ fields $\{\phi_i\}$ to $N-1$ fields constrained to a locus in field space,
such as the ellipse of \cite{shapo}. 

We clearly see that
$\bar{K}$ is the order parameter because the 
decay constant of the dilaton
is precisely $f=\sqrt{2\bar{K}}$, in analogy 
to $f_\pi$ in a chiral Lagrangian, or the VEV of the Higgs
in the standard model. The constraint of constant $\bar{K}(\{\phi_i\})$
gurantees that the dilaton fluctuation is orthogonal in the kinetic
terms to the other 
$N-1$ constrained fields and neatly factorizes.

We emphasize that this is a dynamical process. Just as 
steam can condense into water, a scale disordered phase can condense 
into a scale ordered one.   All of this is tracked in a single frame, which
begins as a Jordan frame.  In this view the universe is a physical system that starts 
in one phase, which has no scale ordering,  and ends up in another in which the scale
SSB defined by $\bar{K}$.  
This is treated in one set of ``frame variables'' with an Friedman-Roberston-Walker metric.
In  a sense the approach of $K\rightarrow \bar{K}$ is just the relaxation of
the dilaton $\sigma\rightarrow 0$, though the dilaton can only be defined in 
the broken phase of the theory.

Here we need not do the Weyl transformation along the way and the 
SSB materializes dynamically.
However, at late times the dynamically generated $\bar{K}$ can then be
matched to the scale quantities, $M_P$ , $\Lambda$, etc., in an Einstein-Hilbert action.
Once these scales are identified, then it is useful
to make a Weyl transformation, e.g., using $\bar{K}$, to isolate the dilaton. 
This guarantees that the dilaton factorizes and alleviates
any putative messy kinetic term mixing issues.
(in fact, this permits the dilaton to be ``eaten'' by a Higgs mechanism if we introduce Weyl's photon, $\hat{A}_\mu$
as in Section II.D, allowing the Weyl photon and dilaton to decouple as very heavy states.)
There does then remain a mixing issue amongst the remaining $N-1$ constrained fields, and these
must be diagonalized to apply the low energy dynamics.

The advantage of phrasing things in terms of conserved currents
is that the results are model independent. 
We never have to actually construct and solve difficult nonlinear
partial differential equations of motion to see this;
this will happen automatically, and the resulting mass scales, including the Planck mass, are generated
spontaneously, controlled by 
the Weyl current.  This mechanism does not depend upon
a potential, (though the particular final vacuum state is dictated by a potential).  
The statements we make are general and model independent,
similar to those of any traditional  ``current algebra.''

A crucial aspect of this mechanism is that quantum theory should not break scale symmetry. 
We believe this
is generally possible. To understand this, it is important that one does not conflate the 
procedure of regularisation, which generally introduces arbitrary mass scales, with renormalisation,
which introduces counter-terms to define the final theory and its symmetries. 
Though it may be convenient, one need not deploy a regulator that is consistent 
with the symmetries of the renormalised theory. The nonexistence of a symmetry in the 
regulator does not imply the nonexistence of the symmetry in the renormalised theory. 
Furthermore, physics should not depend upon the choice of regulator \cite{bardeen}. 

In this view, Weyl symmetry is central and all mass scales must emerge by way of random initial 
conditions governing VEV's (Vacuum Expectation Values) of fields that are entirely contained 
within the action. Essentially there exist no fundamental mass scales, and the mass of anything 
is defined only relative to field VEV's in the theory. For this to be phenomenologically
acceptable it is necessary to explain how the spontaneous breaking of Weyl symmetry can 
lead to a period of inflation followed by a reheat phase and transition in the infrared 
to a theory describing the fundamental states of matter and their interactions with an 
hierarchically large difference between the Planck scale and the electroweak breaking scale. 

Remarkably it has been shown in a simplified model involving two scalar fields that this 
structure is possible \cite{shapo,GB,FHR}. The model has a scale invariant scalar potential 
and non-minimal coupling of the scalar fields to the Ricci scalar. When the fields develop 
VEVs the Planck scale is generated spontaneously in the Brans-Dicke manner. For a wide range 
of the non-minimal couplings and scalar interactions, there is an initial period of ``slow-roll''
inflation that can give acceptable values for the slow-roll parameters. This is followed by a ``reheat" 
phase and a flow of the field VEVs to an infrared fixed point at which the
ratio of the scalar field VEVs are determined by the dimensionless couplings of the theory. 
Thus it is possible to arrange an hierarchically large ratio for the VEVs and, interpreting 
the second scalar as modelling the Standard Model Higgs boson, this large ratio corresponds 
to the ratio between the Planck scale and the electroweak scale.

In section \ref{sinertial} we discuss the mechanism of inertial
spontaneous symmetry breaking and conservation of
the Weyl current in a toy model, and general $N$-scalar models. 
As it does not involve a potential the mechanism 
opens a new pathway to generating spontaneous scale symmetry breaking 
and the associated spontaneous breaking of other symmetries. As such it 
may be useful for novel aspects of model building.  We also discuss a general feature of this
mechanism, the origin of the dilaton and its intimate relationship to the current, 
We also briefly consider, as an aside, locally Weyl invariant models
in which the dilaton will be eaten by a ``Weyl photon,'' $\hat{A}_\mu$, to give it mass,
i.e., the inertial symmetry breaking thus
becomes a Weyl symmetry Higgs mechanism, and the dilaton disappears from
low energy physics \cite{Ferreira:2016kxi}. 

In section \ref{sscale} we discuss how Weyl invariance is maintained at the quantum level 
and thus preserves the inertial spontaneous symmetry breaking mechanism. 
As a result the logarithmic corrections that normally break the 
scale invariance now automatically depend only on physically relevant ratios of field VEVs which preserve
the underlying Weyl invariance of the theory.
We compare this procedure to previous proposals for scale invariant regularisation that require 
an arbitrary choice of regulator, a function of the scalar fields.  

Finally, in section \ref{ssummary}, we present a summary of our results and the conclusions to be drawn.

\section{Inertial spontaneous symmetry breaking.}\label{sinertial}

\subsection{A Toy Example}

Consider a real scalar field theory action together with 
Einstein gravity and a cosmological constant
(our metric signature convention is $(1,-1,-1,-1)$):
\bea
\label{eqone}
S=\int \sqrt{-g}\left(
\frac{1}{2}g^{\mu \nu }\partial_{\mu }\sigma \partial_{\nu}\sigma 
  -\Lambda +\half M_P^2 R \right)  .
\eea
This action provides a caricature of the cosmological world we live in. 

We imagine an initial, ultra-high-temperature phase in which the massless
scalar $\sigma$ has the dominant energy density, $\rho_\sigma \propto T^4$. 
Consider a Friedman-Robertson-Walker (FRW) metric:
\bea
g_{\mu\nu }& =&  [1,-a^{2}(t),-a^{2}(t),-a^{2}(t)] \qquad H=\frac{\dot{a}}{a} \;.
\eea 
In this theory the universe initially 
expands in a FRW phase, with the temperature
red-shifting as $T\sim 1/a(t)$, and the scale factor growing as $a(t)
\sim \sqrt{t}$.  Eventually the $\sigma$ thermal energy becomes smaller
than the cosmological constant, $\rho_\sigma < \Lambda$, and we then enter a deSitter
phase with exponential growth, $a(t)\sim e^{t\sqrt{\Lambda/3M^2_{P}}}$.
We can model the thermal phase as a pre-inflationary era, and the 
cosmological constant then represents a potential energy that drives inflation.
In any case, the intuition that allows us to readily understand how this works 
is well-honed.

Now consider a  different  action:
\bea
\label{eqtwo}
S=\int \sqrt{-g}\left(\frac{1}{2}g^{\mu \nu }\partial_{\mu }\phi \partial _{\nu
}\phi  -\frac{\lambda}{4}\phi^4 -\frac{\alpha}{12}\phi^2 R \right)  .
\eea
This action is scale invariant, having no cosmological constant or
Planck scale.  

These two theories 
are classically equivalent, provided $\alpha<1$. This equivalence 
follows from a Weyl transformation.
But, our question then becomes, given that, from our accumulated experience in 
inflationary cosmology we understand the dynamics of
eq.(\ref{eqone}) so well,  then how
could we directly understand the dynamics of the Weyl equivalent eq.(\ref{eqtwo})
without performing a Weyl transformation into eq.(1)?  
What happens in the pure
evolutionary dynamics intrinsic to eq.(3) that produces the physical mass scales
of $M_P$ and $\Lambda$, as well as all other scales in nature?

At first this doesn't look too hard.  Indeed, if $\phi$ starts out in 
some very high-temperature phase, where the energy density
is large compared to $\lambda \phi^4$
then we expect the scale factor will increase
in a scale invariant way, $a(t)\sim t$. This follows by intuiting that the Hubble constant
satisfies $H^2\sim T^4/\phi^2$, where the $\phi^2$
factor in the denominator replaces $M_P^2$. In thermal equilibrium we expect $\phi^2 \sim T^2$ 
and thus $H=\frac{\dot{a}}{a}\sim T\sim \frac{1}{t}$. Therefore, $a(t)\sim t$.

As the universe cools, we expect ${\phi}(x)$ to settle into some spatially
constant VEV  $\VEV{\phi}$.  However, our intuition from conventional Einstein $M_P^2R$ gravity tells us that
this VEV will slow-roll in the potential, with $\VEV{\phi}$ eventually becoming zero. 
In eq.(\ref{eqtwo})
this would then imply a vanishing $M_P$,   and the details of the
solution become less clear.  It is plausible that the increasing strength of gravity
will increase the Hubble damping, and halt the relaxation of $\VEV{\phi}$,
perhaps leading  to a nonzero cosmological constant $\lambda \VEV{\phi}^4$. 
If true, this would then match the 
cosmological constant case of eq.(\ref{eqone}), and it would imply a spontaneous
breaking of scale symmetry.  We could resort to a numerical solution,
but how can we see what happens in 
a simple and intuitive way, without having to puzzle over the solutions
of coupled nonlinear differential equations?

Indeed, from eq.(\ref{eqtwo}) we can directly obtain the Einstein equation:
\bea 
\label{einstein1}
 \!\!
\frac{1}{6}\alpha \phi ^{2}G_{\alpha \beta } & = &
\left(\frac{3-\alpha}{3} \right)\partial _{\alpha }\phi \partial _{\beta }\phi
-g_{\alpha \beta }\left( \frac{3-2\alpha}{6}\right) \partial
^{\mu }\phi \partial _{\mu }\phi 
\nonumber \\
& & \!\!\!\!\!  \!\!\!\!\! \!\!\!\!\! \!\!\!\!\!     
+\frac{1}{3}\alpha \left( g_{\alpha \beta
}\phi D^{2}\phi -\phi D_{\beta }D_{\alpha }\phi \right) +g_{\alpha \beta
}V(\phi) .
\eea
The trace of the Einstein equation becomes:
\bea 
\label{trace1}
\!\!\!\!\!\!\!\!\!\! \!\!\!\!\!\!
-\frac{1}{6}\alpha \phi ^{2}R & = &  (\alpha-1 )\partial ^{\mu }\phi
\partial _{\mu }\phi +\alpha \phi D^{2}\phi +4V(\phi).
\eea
We also have the Klein-Gordon (KG) equation for $\phi$:
\bea
\label{KG1}
0=\phi D^{2}\phi+\phi\frac{\delta }{\delta \phi }V\left( \phi
\right) +\frac{1}{6}\alpha \phi^2 R.
\eea
We can combine the KG equation, eq.(\ref{KG1}), and trace equation, 
eq.(\ref{trace1}), to eliminate the $ \alpha \phi^2 R $ term, and obtain:
\bea
\label{Kdiv00}
0& =& (1-\alpha )\phi D^{2}\phi +(1-\alpha )\partial ^{\mu }\phi \partial
_{\mu }\phi 
\nonumber \\
& & \qquad\qquad +\phi \frac{\delta }{\delta \phi }V\left( \phi \right) -4V\left(
\phi \right) .
\eea
This can be written as a current divergence equation:
\bea
\label{Kdiv1}
D^\mu K_\mu = 4V\left(
\phi \right) -\phi \frac{\partial }{\partial \phi }V\left( \phi \right) .
\eea
where 
\bea
K_\mu=(1-\alpha)\phi\partial_\mu \phi
\eea
is the ``Weyl current.''
For the scale invariant potential, $V(\phi)\propto \phi^4$, the {\em rhs}
of eq.(\ref{Kdiv1}) vanishes
and the $K_\mu$ current is then covariantly conserved:
\bea
\label{cons1}
D^\mu K_\mu = 0.
\eea
We see that this is an ``on-shell'' conservation law, i.e., 
it assumes that the gravity satisfies eq.(\ref{einstein1}).
This is the global Weyl current and it can be derived by a Noether variation
of the action under a Weyl transformation.

Note that the Weyl current, $K_\mu$, 
is the derivative of a scalar,  $K_\mu =\partial_\mu K$, 
where:
\bea
K=\frac{1}{2}(1-\alpha)\phi^2.
\eea
We refer to $K$ as the ``kernel.''
Using the conserved $K$-current with its kernel, we can easily understand the dynamics of this 
theory. 

The form of the conservation law is
$D^\mu K_\mu = D^2K=0$, and this holds in any frame. If we take 
$\phi$ to be functions of time $t$ only, and consider a Friedman-Robertson-Walker universe
($g_{\mu \nu }=[1,-a^{2}(t),-a^{2}(t),-a^{2}(t)]$) 
the current conservation equation implies:
\begin{eqnarray}
{\ddot { K}}+3\left(\frac{\dot a}{a}\right){\dot { K}}=0.
\end{eqnarray}
This can be readily solved to give:
\begin{eqnarray}
K(t)=c_1 + c_2\int_{t_0}^t \frac{dt'}{a^3(t')},
\end{eqnarray}
where $c_{1,2}$ are constants. Therefore in an expanding universe {\em $K$
will evolve to a constant value,} ${K}\rightarrow \bar{K}$. 

In the single scalar case, as $K\rightarrow \bar{K}$ constant, the initial Jordan frame theory 
flows to an effective final Einstein-Hilbert theory with parameters
$\Lambda =\frac{\lambda {\bar{K}}^2}{(1-\alpha)^2},\;M_P^{2}=-\frac{\alpha{\bar{K}}}{3(1-\alpha)},  
\;\; f^2=2{\bar{K}}$ (dilaton decay constant, see II.B) \cite{FHR}. The 
equivalence between the theories is achieved
dynamically, without having performed a Weyl transformation, and it
follows from the Weyl current algebra, and does not rely upon the solutions
of complicated nonlinear differential equations of motion.

This is robust. If we consider a set  of N scalars, $\{\phi_j\}$, with action given 
by:\footnote{It is straightforward to extend this effective Lagrangian to matter 
and gauge fields
\cite{shapo,Ferreira:2016kxi,UW}.}
\begin{eqnarray} \label{msaction}
S &=& \int d^4 x\sqrt{-g}\left[\frac{1}{2}\sum_{i}^{N}\partial_\mu\phi_i\partial^\mu\phi_i  -W(\{\phi_j\})
\right.
\nonumber \\
& & \qquad \qquad\qquad  \left.-\frac{1}{12}F(\{\phi_j\})R \right].  
\end{eqnarray}
where we maintain scale invariance (i.e., $F(\{\phi_k\})$ and $W(\{\phi_k\})$ transform respectively  as
$F\rightarrow e^{2\epsilon}F$ and $W\rightarrow e^{4\epsilon}W$  under 
global Weyl transformations, as defined below in
 eq.(\ref{transforms})).
The conserved Noether current kernel then generalises to:
\begin{eqnarray}
K=\frac{1}{2}\left[\left(\sum_{i=1}^N\phi^2_i\right) -F(\{\phi_k\})\right]
\label{locus}.
\end{eqnarray}
In particular, with $F(\{\phi_j\})=\sum_{i}^{N}\alpha_i\phi_i^2$
the kernel takes the form \cite{FHR}:
\begin{eqnarray}
K=\frac{1}{2}\sum_{i=1}^N(1-\alpha_i)\phi^2_i 
\label{ellipse}.
\end{eqnarray}
In this case the $N$ scalar fields will evolve such that 
their values will ultimately be constrained to lie on the  $N$-dimensional
locus by eq.(\ref{locus}) with $K\rightarrow \bar{K}$, in particular
an ellipsoid in the special case of eq.(\ref{ellipse}).

Here we are ``launching'' the theory 
in an effective Jordan frame, with arbitrary initial
values of the fields and their time derivatives $\{\phi_i, \dot{\phi_j}\}$.
The initial expansion will be scale invariant, $a(t)\sim t$, but as $K\rightarrow \bar{K}$,
the Planck scale becomes dynamically established, and 
we enter an effective Einstein frame where all mass scales are $\propto \sqrt{\bar{K}}$, and
the expansion becomes deSitter, $a(t)\sim \exp\sqrt{\bar{K}}t)$.

In a 2-scalar model discussed in ref.\cite{FHR} we
have checked numerically that the initial rate of approach to 
the ellipsoid is rapid and thereafter the fields precisely 
track the ellipsoid corresponding to constant $\bar{K}$. 
This is true for a wide range of initial conditions and readily 
allows for an inflationary period to commence.
Since $K$ has dimension of $($mass$)^2$, a constant vacuum value of $K$ implies
a spontaneous breaking of the scale symmetry in the theory has occurred. 
Note that this phase does not employ 
a potential but is driven solely by the initial conditions, and  
$K$ is the order parameter of
inertial spontaneous symmetry breaking.

In multi--scalar theories the flow $K\rightarrow \bar{K}$ does not fix the relative values 
of the scalar field VEVs, which initially end up at some random point
on the locus (e.g. ellipse). It is here that the potential becomes important. 
In the infrared (IR), the fields constrained to the locus,
flow towards an IR fixed point in which the 
ratios of the field VEVs are determined by the potential terms alone  \cite{FHR}. 
For the case that the potential has a flat direction, 
the vacuum energy vanishes at the minimum, 
corresponding to vanishing cosmological constant.  The IR fixed point
is then the intersection of the potential's flat direction with the locus.
The ratios of the VEV's is then
determined by the scalar potential couplings, but constrained by the requirement 
the fields lie on the $N$-dimensonal ellipsoid.

For the case that the potential 
is positive definite, the IR fixed point corresponds to an eternally inflating 
de-Sitter solution in which the ratio of the field VEV's is determined by the 
scalar potential couplings together with the couplings, $\alpha_i$, of the scalars 
to the Ricci scalar.

\subsection{General Discussion}

Inertial spontaneous symmetry breaking can be 
responsible for triggering the spontaneous breaking of symmetry in all sectors of the theory.
As such it opens new possibilities 
for model building.

In summary, we found that the expansion of the 
universe in a pre-inflationary
phase drives  the current charge density, $K_0$, to zero. 
The global Weyl current, $K_\mu$, 
is {\em always} the derivative of a scalar,  $K_\mu =\partial_\mu K$, 
and in particular $K_0 =\partial_0 K$, where $K$ is the kernel.
Hence, as the $K_\mu$ current density is diluted away, 
$K_0\rightarrow 0$,  the kernel $K$ therefore evolves as   $K\rightarrow \bar{K}$ constant.  
In a Weyl invariant theory this implies that scale symmetry is broken,
and the Planck mass
is generated dynamically.  

$K$ plays the role of the symmetry breaking order
parameter. 
While a potential may then be needed to engineer the
final vacuum, and determine the ratios of individual
fields $\VEV{\phi_i}$, it plays no direct role 
in the inertial Weyl symmetry breaking phenomenon.

With a little
thought,
one might have guessed the structure of the order parameter $K$.
Consider a set of $N$ scalar fields $\{\phi_i\}$.
If the fields are non-minimally coupled
to gravity as $(-1/12) \sum_{i} \alpha_i\phi_i^2 R(g)$,
then if any of the $\phi_i$ should develop a VEV,
we would expect scale breaking, and a nonzero $K$. Hence,
we expect that the order parameter takes the form, $K\sim c\sum_i \phi_i^2$.
However, if any  $\phi_i$ has $\alpha_i = 1$, then we can
remove it from the action by a local Weyl transformation, absorbing it into the metric.
We therefore expect $K= c'\sum_i (1-\alpha_i) \phi_i^2$.
Indeed, we found that $K_\mu = \partial_\mu K$,  with $c'=1/2$,  combining
both the trace of the Einstein and KG equations, or
by the Noether variation
of the Jordan frame theory under a Weyl transformation, thus confirming our guess.

\subsection{Factorization of the Dilaton}

We've seen the  result that $K\rightarrow\bar{K}$ constant as the universe expands
implies that $N-1$ fields $\{\phi'_i\}$  will ultimately
satisfy a constraint, such as in eq.(\ref{ellipse}),
$\bar{K}=(1/2)\sum_i (1-\alpha_i) \phi'_i{}^2$.
Here the constrained fields, $\{\phi'_i\}$, lie on an ellipsoid in field space,
but the constraint could be more general as in eq.(\ref{locus}) with $F(\{\phi_i\})$, 
and the ellipsoid could be a more general locus in field space.

In any case, there remains one field unconstrained 
that becomes the dilaton. This is intimately
related to the $K_\mu$ current. Let us perform a 
Weyl field redefinition on the $N$ original fields,
\bea
\phi_i = \exp(\sigma/f)\phi_i'  \qquad
 g_{\mu\nu}= \exp(-2\sigma/f) g'_{\mu\nu}.  
\eea
We thus find the Weyl invariant action becomes:
\bea
S(\phi, g) & =& S(\phi', g') \nonumber \\
& & \!\!\!\!\!  \!\!\!\!\!  \!\!\!\!\!   \!\!\!\!\! \!\!\!\!\!\!\!\!\!\!
+ \int \sqrt{-g'} \left( \partial_\mu \bar{K}(\phi')\partial^\mu (\sigma/f) 
+ \bar{K}(\phi') (\partial \sigma/f)^2 \right)
\eea
Now using the constraint that $\bar{K}$ constant, and integrating by parts,
we have:
\bea
S(\phi, g) = S(\phi_i', g') 
+ \frac{1}{2}\int \sqrt{-g'} (\partial \sigma)^2 
\eea
Here we identify $f^2=2\bar{K}$ so the dilaton is
canonically normalized.
From this we see that the dilaton, $\sigma$, describes a dilation
of the ellipse, and fluctuates in field space orthogonally to the $N-1$  $\{\phi'_i\}$ fields.
The dilaton decouples in the action
from everything except gravity (this holds true for fermions and gauge bosons
as well; decoupling implies that there are no direct couplings {\em in the action}
to other fields).

This result is elegantly simple. There is no messy kinetic term mixing problem
of the dilaton with the remaining $\phi'$ fields, as some authors have alluded to.
Indeed, there is nontrivial mixing amongst the $\phi'$ that are subject to the constraint,
but the dilaton is neatly factorized and does not mix with these other fields kinetically.

We further see that the current written in the unconstrained
fields is equivalent to one written in the constrained fields
by: $K_\mu = \partial_\mu K(\phi)= \partial_\mu (K(\phi') e^{2\sigma/f})$.
Hence in the broken phase (Einstein frame) limit $K(\phi')\rightarrow \bar{K}$ constant, 
$K_\mu \rightarrow 2\bar{K}\partial_\mu \sigma /f=f \partial_\mu \sigma$.
where $f = \sqrt{2\bar{K}}$.
This is as  we expect for a Nambu-Goldstone boson, e.g., the axial current of the pion takes 
the analogous form $f_\pi \partial_\mu \pi$.  This implies that $\bar{K}$
is the order parameter of Weyl spontaneous symmetry breaking.

Why is this formulation important?
Results following from the ``current algebra''
of Weyl invariant theories are general statements that are true,
independent of the specific structure of the Lagrangian. The particular structure
of $K_\mu$ and $K$ is independent of the form
of any scale invariant potential, but the detailed structure of $K$ does
depend upon the choice of the non-minimal
couplings, e.g., $F(\phi_i)$ in eq.(\ref{locus}) (and also any higher 
derivative gravitational terms can modify the
simple forms we just discussed). The behavior of the current algebra will remain intact,
since $K_\mu=\partial_\mu K$ is conserved, 
but the constraint defined by $\bar{K}$ could become a more general locus
such as a hyperbola, etc., (such effects result from the renormalization
group \cite{FHR}). 
The survival of the general feature of inertial breaking with
a stable goundstate, e.g., a stable $M_{Planck}$, requires that the
quantum theory does not break Weyl symmetry through loops,
as we discuss in Section III. 

\subsection{Local vs. Global Weyl invariance; Eating the Dilaton}

Our main discussion is based upon globally Weyl invariant theories. However,
we include the present section to indicate how it may be possible to promote
these to locally Weyl invariant theories by introduction of Weyl's gauge field,
i.e., ``Weyl's photon.''  It is interesting that inertial symmetry breaking now becomes a Higgs mechanism,
since the Weyl photon will ``eat'' the massless dilaton and thus remove it from
the low energy spectrum, where it becomes the longtudinal degree of
freedom of a massive Weyl photon. Hence, in this case the issue of long range 5th force
limits becomes moot.  The present section is classical, but
it would be of interest to develop the full quantum (renormalization group)
 behavior of Weyl's photon.

Weyl's original idea was that, since coordinates are merely
numbers invented by humans to account for events in space-time, they
should not carry length scale \cite{Weyl}.  Rather, the concept  of length
should be relegated
to the (covariant) metric, and (contravariant) coordinate differentials are scale free.
Therefore, under a local Weyl scale transformation we would have:
\bea
\label{transforms}
g_{\mu\nu}(x) \rightarrow e^{-2\epsilon(x)}g_{\mu\nu}(x) \qquad
&& g^{\mu\nu}(x)\rightarrow e^{2\epsilon(x)}g^{\mu\nu}(x) \nonumber \\
\sqrt{-g}\rightarrow e^{-4\epsilon(x)}\sqrt{-g}\qquad 
&&\phi(x)\rightarrow e^{\epsilon(x)}\phi(x)
\eea
Weyl transformations are distinct from coordinate diffeomorphisms that 
define scale transformations on coordinates,
as $\delta x^\mu=\epsilon(x) x^\mu$, which we discuss below.  
The global Weyl symmetry 
corresponds as usual to $\epsilon=$(constant in spacetime).  

It is straightforward to construct a list of local Weyl invariants:
\bea
\!\! \!\!\phi^2(x)g_{\mu\nu}(x); \!\! \! && \phi^{-2}(x)g^{\mu\nu}(x); 
\;\;\;
\sqrt{-g}(x)\phi^4(x); \nonumber 
\\
R(\phi^2g_{\mu\nu})&=&\phi^{-2}R(g_{\mu\nu})+6\phi^{-3}D^\mu(\partial_\mu\phi) \nonumber
\\
\label{relation}
\sqrt{-g}\phi^4R(\phi^2g_{\mu\nu})&=&\sqrt{-g}\left(\phi^2R(g_{\mu\nu})
+6\phi D^\mu(\partial_\mu\phi)\right)\nonumber
\\ ...
\eea
Note that the computation of $R(\phi^2g_{\mu\nu}) $ above
requires that any Christoffel symbols used in the definition of $R$ be evaluated 
in the metric $\phi^2 g_{\mu\nu}$.
Using these identities we can construct an action that is locally
Weyl invariant:
\bea
S&=&\int d^4x\;\sqrt{-g}\left(-\frac{1}{12}\phi^4R(\phi^2g)-\frac{\lambda}{4}\phi^4 \right) \\
&=&\int d^4x\;\sqrt{-g}
\left(\frac{1}{2}\partial_\mu\phi\partial^\mu\phi-\frac{1}{12}\phi^2 R(g)-\frac{\lambda}{4}\phi^4 \right)  
\nonumber 
\eea
where we substituted the relationship of eq.(\ref{relation}) and integrated by parts using the divergence
rule $D_\mu V^\mu= \sqrt{-g}^{-1}\partial_\mu (\sqrt{-g}V^\mu)$. Here we obtain the
famous locally Weyl invariant theory in which the nonminimal coupling of scalars to gravity
is fixed by the coefficient $1/12$, needed to canonically normalize the
$\phi$ kinetic term.  This is a special and somewhat degenerate
theory, since we
can revert to the metric $\hat{g}_{\mu\nu}=\phi^2 g_{\mu\nu}$ and $\phi$ disappears
from the action.
The theory has a vanishing Weyl current \cite{JP}.

We note that covariant gauge fields, 
such as the electromagnetic vector potential, $A_\mu $,
do not transform under the local Weyl transformation,  since they are associated with 
 derivatives $\partial_\mu-ie A_\mu $ which, like coordinates, do not transform. The 
 electromagnetic fields that have the usual engineering scale 
 $\sim$(mass)$^2$, $\vec{E}$ and $\vec{B}$, are contained in
 the field strength with one covariant and one contravariant index,
 $F_\mu^{\;\;\nu}$, e.g., $\vec{E}_i = F_i^0$. 

We can construct
a covariant derivative of a scalar field under local Weyl transformations by
introducing the ``Weyl photon,''  $\tilde{A}_\mu$, as 
\bea
\tilde D_\mu\phi=\partial_\mu\phi -\tilde{A}_\mu\phi
\eea
where $\phi(x)\rightarrow e^{\epsilon(x)} \phi(x)$ and 
$\tilde{A}_\mu(x)\rightarrow \tilde{A}_\mu(x) +\partial_\mu \epsilon(x)$  
(note the major difference from electrodynamics in the absence of a factor of $i$
in the coefficient of $\tilde{A}_\mu$: QED gauges phase, while the Weyl photon gauges scale).
Armed with this we can construct another local Weyl invariant:
\bea
\label{KTgauge}
\sqrt{-g} g^{\mu\nu} \tilde D_\mu\phi(x) \tilde D_\nu\phi(x).
\eea
This is a locally Weyl invariant kinetic term.\footnote{ 
Here there is a subtlety, as we must define the derivative of any
conformal field as a commutator: $[D_\mu, \Phi] =\partial_\mu\Phi -A_\mu[W,\Phi]$
where $[W,\phi] = w\phi$ and $w$ is the conformal charge of $\Phi$.
Hence $w=1$ for $\phi$. We also require $w=-2$ for $g_{\mu\nu}$,
$w=+2$ for $g^{\mu\nu}$, $w=-4$ for $\det{-g}$, etc. Note that
$[D_\mu, g_{\rho\sigma}] =D_\mu g_{\rho\sigma} +2\tilde A_\mu g_{\rho\sigma}$
$=\tilde A_\mu g_{\rho\sigma}$ since $D_\mu g_{\rho\sigma}=0$.
This insures
the invariance of the action with the Weyl covariant derivative 
under integration by parts.
Note that we can alternatively define a restricted
``pure gauge theory'' with $A_\mu =\partial_\mu \ln(\chi)$,
where $\chi$
is any massless scalar field.}

We can combine this
with the previous invariants to define an action in which the
Weyl symmetry is local, yet the
nonminimal coupling of scalars to $R$ is arbitrary:
\bea
\label{local}
S&=&\int d^4x\;\sqrt{-g}\left(
 \half (1-\alpha) g^{\mu\nu}\tilde D_\mu\phi \tilde D_\nu\phi
-\frac{\lambda }{4}\phi^4 \right. \nonumber \\
& &\qquad  \qquad \qquad \left.
-\frac{\alpha}{12}\phi^4R(\phi^2g)\right)\nonumber \\
&=&\int d^4x\;\sqrt{-g}
\left(
\frac{1}{2}g^{\mu\nu}\partial_\mu\phi\partial_\nu\phi-\frac{\alpha}{12}\phi^2 R(g)
-\frac{\lambda}{4}\phi^4
\right.
\nonumber \\
&& \qquad \left.
- \half (1-\alpha)\left(\tilde{A}^\mu  \partial_\mu(\phi^2) -
\tilde{A}^\mu\tilde{A}_\mu \phi^2 \right) \right)  .
\eea

Now, we want to pass to the Weyl broken phase. We write:
\bea
\phi (x)&\rightarrow & f\exp (\sigma (x)/f)\nonumber \\
g_{\mu \nu }(x)&\rightarrow & \exp (-2\sigma (x)/f)g_{\mu \nu }(x)
\eea
Note we do not at this stage do
a gauge tranformation, $A_{\mu }\rightarrow A_{\mu }+\partial _{\mu }\sigma (x)/f$.
We obtain,
\bea
S&=&\int \sqrt{-g}\left( (1-\alpha )\frac{1}{2}g^{\mu \upsilon }\partial
_{\mu }\sigma \partial _{\nu }\sigma -(1-\alpha )g^{\mu \upsilon }A_{\mu
}f\partial _{\nu }\sigma \right.
\nonumber\\
&& \left. +\frac{1}{2}(1-\alpha )g^{\mu \upsilon }A_{\mu
}A_{\nu }f^{2}-\frac{1}{12}\alpha f^{2}R\right) 
\eea
Note that the Weyl transformation
cancelled the original $\frac{1}{2}\alpha g^{\mu \upsilon
}\partial _{\mu }\phi \partial _{\nu }\phi $ piece since it was local.
What is left is a perfect square;
\bea
S&=&\int \sqrt{-g}\left( \frac{1}{2}f^{2}(1-\alpha )g^{\mu \upsilon
}\left( A_{\mu }-\partial _{\mu }\sigma /f\right)^2 
-\frac{1}{12}\alpha f^{2}R\right) 
\nonumber \\ &=&\int \sqrt{-g}\left( \frac{1}{2}f^{2}(1-\alpha )g^{\mu \upsilon }B_{\mu
}B_{\nu }-\frac{1}{12}\alpha f^{2}R\right) 
\eea
where we redefine $B_{\mu }=A_{\mu }-\partial _{\mu }\sigma /f$
which is a massive spin one field of mass $m=f\sqrt{(1-\alpha )}%
.$  The dilaton has been eaten by the Weyl photon to become 
its longitudinal mode,
and the massless dilaton has thus disappeared 
from the spectrum of the theory.

We can always have a kinetic term
for $A_{\mu }$ with 
\beq
F_{\mu \nu }=-
[D_{\mu },D_{\nu }]=\partial _{\mu }A_{\nu }-\partial _{\nu }A_{\mu
}=\partial _{\mu }B_{\nu }-\partial _{\nu }B_{\mu }
\eeq
and
\beq
S=\int \sqrt{-g}\left( -\frac{1}{4}F_{\mu \nu }F^{\mu \nu }+\frac{1}{2}%
m^{2}g^{\mu \upsilon }B_{\mu }B_{\nu }-\frac{1}{12}\alpha f^{2}R\right) 
\eeq
The equation of motion for $B_{\mu }$ is
\beq
\partial _{\mu }F^{\mu \nu }=D^{2}B^{\nu }-\partial ^{\nu }\left( D_{\mu
}B^{\mu }\right) =m^{2}B^{\nu }
\eeq
This is mathematically analogous to
a superconductor or the Standard model Higgs mechanism.
A gas of $B_{\mu }$ will freeze out and redshift
away like matter once the temperature redshifts
below $m$. It is also interesting to note that
if we have $N$ $\phi_i$ fields, the inertial symmetry breaking
will yield the $N-1$ $\phi'_i$ fields and the dilaton which
is again eaten to become the longitudinal component of $B^{\mu }$,
but we then find that the gauge field $B^{\mu }$ decouples 
from the $\phi'_i$!  It also has even charge conjugation and
presumably decouples from fermions and gauge fields as well, 
and it cannot decay to a pair of gravitons (this is a variation
on Yang's theorem which forbids decay of a vector
meson to a photon pair).
Therefore, relic $B^{\mu }$ fields are stable and could
constitute a dark matter candidate if they are not inflated
away.

From the action of eq.(\ref{local}) we see that the Weyl current is
easily obtained:
\bea
K_\mu & =& -\frac{1}{\sqrt{-g}}\frac{\delta S}{\delta \tilde A^\mu}
= (1-\alpha)\left( \phi\partial_\mu\phi-\tilde{A}_\mu\phi^2\right)
\nonumber \\
&=& (1-\alpha)\phi\tilde{D}_\mu\phi .
\eea
This still has the general form $K_\mu=D_\mu K$, where $D_\mu$ is a covariant Weyl derivative.

By setting $\tilde{A}_\mu=0$ we obtain
a globally invariant theory, and this current
becomes the conserved Noether current for
the global Weyl invariant theory:
\bea
\label{current}
K_\mu 
= (1-\alpha) \phi\partial_\mu\phi.
\eea

 \section{Quantum scale invariance and regularisation }\label{sscale}
 
 Up to now our discussion has been confined to the classical action. 
 For the scenario of inertial spontaneously broken scale symmetry to work, 
 and lead to a stable Planck mass, it is essential the that Weyl current be 
 identically conserved at the quantum level \cite{Englert:1976ep} : 
 \be
 D^{\mu}K_{\mu}=0.
 \label{conserved}
 \ee
 In what follows we will refer to nonzero contributions coming from loops to the
rhs of eq.(\ref{conserved}) as ``Weyl anomalies."  The 
trace anomalies of the scale current determined by diffeomorphisms are 
identical to those in $K$ for the scalar sector of the theory.
 
Scale and Weyl symmetry of a theory appears ab initio to be broken by quantum loops. 
Loop divergences
are subtle, however, and are often confused with physics. Here we adopt 
an operating principle that has been espoused
by W. Bardeen \cite{bardeen}: The allowed symmetries of a renormalised quantum field 
heory are determined by anomalies, (or absence thereof). Quantum loop divergences 
are essentially unphysical artefacts of the method of calculation.
 
 Weyl or scale symmetry is permitted if the renormalised theory has no Weyl anomalies. 
 Since trace anomalies come from triangle diagrams they are necessarily associated 
 with dimension-4 operators. Hence there is no Weyl anomaly in the 
Standard Model of the form $H^\dagger H$ where the Higgs mass is $m^2 H^\dagger H$. 
Thus there are no Weyl anomalies associated with quadratic or quartic divergences 
in quantum field theory in four dimensions. Another way of saying this is that 
divergent terms and counter terms are not separately measurable, only the 
renormalised mass is physical. In a variation of the Standard Model with no gravity, 
no grand unification and no Landau poles in the far UV the Higgs mass would be technically 
natural with no hierarchy problem!

\subsection{The origin of Weyl anomalies}

Our problem of maintaining Weyl symmetry requires that we build a theory that has no anomaly in $K_{\mu}$. 
To understand this problem, and its solution, we turn to the CW potential. In computing CW potentials
for massless scalar fields we encounter an infrared divergence that must be regularised \cite{Coleman:1973jx, CTH}.
To do so we often introduce explicit
``external" mass scales into the theory by hand. These are mass scales that are not part of the defining action of
the theory, and essentially define the RG trajectories of coupling constants. These externally injected mass scales
lead directly to the Weyl anomaly. 

We can see this in  eq.(3.7) of CW \cite{Coleman:1973jx} where, 
to renormalise the quartic scalar coupling constant, $\lambda$, 
in an effective potential at one loop level, $W(\phi)$, they introduce 
a mass scale $M$. Once one injects $M$ into the theory, one has broken scale and Weyl symmetry, 
and the effective potential in the large ${\phi\over M}$ limit then takes the form
\be
W(\phi)={\beta_1\over 4!} \phi^4 \ln\left({\phi\over M}\right)
\label{CW}
\ee
Here ${\beta_1}$ is the one-loop renormalisation group coefficient, $d\lambda(\mu)/ d\mu=\beta_1$.
The manifestation of this is seen in the trace of the improved stress tensor [13], and in the 
divergence of the $K_{\mu}$ current:
\bea
\label{Kdiv3}
\partial^\mu K_\mu = 4W\left(
\phi \right) -\phi \frac{\delta }{\delta \phi }W\left( \phi \right) =-\frac{\beta_1}{4!}\phi^4
\eea
Of course, there is nothing wrong with the CW potential, or with this procedure, if one is only treating the effective potential as a subsector of the larger theory. If, however, Weyl symmetry is to be maintained as an exact invariance of the world, then $M$ must be replaced 
by an internal mass scale that is part of action, i.e. $M$ must then be 
the VEV of a field, $\chi$, or some combination of the fields, 
appearing in the extended action. We would then have the Coleman-Weinberg potential:
\be
W(\phi,\chi)={\beta_1\over 4!} \phi^4 \ln\left({\phi\over \chi}\right)
\label{Vscale}
\ee
and, because we now have no external mass scales, the current divergence vanishes:
\bea
\label{Kdiv2}
\partial^\mu K_\mu &=& 4W\left(
\phi ,\chi\right) -\phi \frac{\delta W\left( \phi ,\chi\right) }{\delta \phi } - \chi \frac{\delta W\left( \phi ,\chi\right) }{\delta \chi}=0.
\nonumber \\
\eea
This defines the basic idea for maintaining scale symmetry in the quantum theory.
It simply implements the notion that there are no fundamental mass scales,
and masses are determined only as dimensionless ratios involving VEV's of scalar fields.
In the next section we illustrate this through a calculation of the one loop correction to the scalar potential arising from the quartic scalar interaction. Of course there will be further gravitational corrections but their calculation lies beyond the scope of this paper.

 \subsection{Weyl Invariant Coleman-Weinberg Calculation}
 
 How might we derive such a result as in eq.(\ref{Vscale}) from first 
 principles? We do so via a computation
 of a Coleman-Weinberg (CW) effective potential. It is important to realise that CW effective potentials 
 themselves must have the full symmetry of the underlying theory. 
 The symmetry is then broken spontaneously by the minimum of the potential. 
 
In fact it is straightforward to show that the usual regularisation 
procedure applied to the Weyl invariant theory of eq.(\ref{msaction}) {\it does} 
have a Weyl invariant form. For the simple two scalar case, $N=2$, 
with fields $\phi=\phi_1$ and $\chi=\phi_2$,  
it reduces to that of eq.(\ref{Vscale}) when the ratio of VEV's is small, but the general 
form is applicable for arbitrary values of the ratio.

\subsubsection{The two scalar action}

The case, $N=2$, is the simplest model with ``realistic" phenomenological properties. 
For reasonable parameter choices and initial conditions it can have an initial inflationary 
period followed by a ``reheat" phase and subsequent evolution to an IR stable fixed point in 
which the ratio of the field VEVs is determined by the fudamental couplings of the theory. 
We will illustrate the regularisation procedure applied to this model (in the limit
of neglecting graviton loops) but we emphasise that the 
procedure immediately generalises to the case with arbitrary $N$ and indeed to the inclusion 
of fundamental fermions and vectors.

We start with the action given in eq.(\ref{msaction}) with $N=2$. 
The Weyl invariance of the theory is spontaneously broken by the VEVs 
of the fields giving a massless Goldstone boson, the dilaton, $\sigma$. 
It was shown in \cite{Ferreira:2016kxi} that the dilaton decouples and 
so, of the two initial scalar degrees of freedom, only one interacting 
one remains. To see how this happens in practice, we change variables to:
\begin{eqnarray}
\label{W11}
\phi_i&=& e^{-{\sigma}/{f}} {\hat \phi}_i\nonumber \\
g_{\mu\nu}&=&e^{2{\sigma}/{f}} {\hat g}_{\mu\nu}
\end{eqnarray}
where ${\hat \phi}_i$ are constrained to lie on the ellipse given by:
\begin{eqnarray}
2{\bar K}=\sum_{i=1}^N(1-\alpha_i){\hat \phi}^2_i=f^2
\end{eqnarray}
where $f^2$ is a constant.
 It is important to note that $f$ is invariant under scale transformations as the dilaton dependence of the original fields has been factored out. 
 
 To illustrate the regularisation procedure it is sufficient to calculate the 
 CW potential resulting from the ${\lambda\over 4!}\phi_1^4$ term in the potential.
 We first re-parameterise the fields by:
\be
\label{constrained}
\hat\phi_1={f\over \sqrt{1-\alpha_1}}\;\sin\theta,\;\;\hat\phi_2={f\over \sqrt{1-\alpha_2}}\;\cos\theta
\ee
After scaling out the dilaton,  the relevant terms of eq.(\ref{msaction}) become:
\bea
S& =&\int d^4x \sqrt{-\hat g}
\left[{1\over 2}f^2\left( {\cos^2\theta\over (1-\alpha_1)}+{\sin^2\theta\over (1-\alpha_2)}\right)\partial_{\mu}
\theta\partial^{\mu}\theta \right.
\nonumber \\
& & \left. \qquad\qquad \qquad
-{\lambda\over 4} f^4{\sin^4\theta\over (1-\alpha_1)^2}   \right ]
\label{2scalarS}
\eea
Performing the further redefinition $\Theta=F(\theta)$ where:
\be 
F(\theta)=\int^\theta_0\sqrt{ {\cos^2\theta'\over (1-\alpha_1)}+{\sin^2\theta'\over (1-\alpha_2)}}\;d\theta'
\ee
the action becomes: 
\be
S=\int d^4x \sqrt{-\hat g}\left[{1\over 2}f^2\partial_{\mu}\Theta\partial^{\mu}\Theta -{\lambda\over 4!} f^4{\sin^4F^{-1}(\Theta)\over (1-\alpha_1)^2}   \right ].
\ee
For the case $\theta$ is small the action approximates to the simpler form:
\be
S\approx\int d^4x \sqrt{-\hat g}\left[{1\over 2}\partial_{\mu}\Phi\partial^{\mu}\Phi -{\lambda\over 4!} \Phi^4   \right ].
\label{act1}
\ee
where  $\Phi=f \Theta$ and $\Theta\approx{\theta\over\sqrt{1-\alpha_1}}$.

\subsubsection{The CW potential}

Here we demonstrate the derivation of the Weyl invariant CW potential for the case 
${\phi_1\over \phi_2}\ll 1$, starting with the action of eq.(\ref{act1}). 
Adding a classical source term, $-J\Phi$, to the Lagrangian induces a shift in the $\Phi$ field:
\be
\Phi=\Phi_c+\hbar^{1/2}\hat \Phi
\ee
where $\hat \Phi$ is the small fluctuation about the classical minimum.
 Thus the potential has the form:
\be
 W(\Phi) = \frac{\lambda }{{4!}}\Phi _c^4 + \hbar \frac{\lambda}{4}\Phi _c^2{\widehat \Phi ^2} + ...
 \ee
 where the linear term cancels due to the classical source term. 
 Treating the quadratic term in $\hat\Phi$ as an interaction the 1-loop potential 
 with $\hat\Phi$ the propagating field is given by:
\bea
W_{eff}&=&  \Omega
 + i\int {\frac{{{d^4}k}}{{{{\left( {2\pi } \right)}^4}}}\sum\limits_{n = 1}^\infty  {\frac{1}{{2n}}{{\left( {\frac{{{\textstyle{1 \over 2}}\lambda \Phi _c^2}}{{{k^2} + i\varepsilon }}} \right)}^n}} } \nonumber 
\\
 &=& \Omega + \frac{1}{2}\int {\frac{{{d^4}k}}{{{{\left( {2\pi } \right)}^4}}}
 \ln \left( {1 + \frac{{\lambda \Phi _c^2}}{{2{k^2}}}} \right)} \nonumber \\
 &=& \Omega
+ \frac{{\lambda {\Lambda ^2}}}{{128{\pi ^2}}}\Phi _c^2 
- \frac{{{\lambda ^2}\Phi _c^4}}{{256{\pi ^2}}}
\ln \left( {\frac{{{\textstyle{1 \over 2}}\lambda \Phi _c^2+{\Lambda ^2}}}{{{\textstyle{1 \over 2}}\lambda \Phi _c^2}}} \right) 
\nonumber \\
& & \qquad+ \frac{{{\Lambda ^4}}}{{64{\pi ^2}}}\ln \left( {\frac{{{\textstyle{1 \over 2}}\lambda \Phi _c^2 
+ {\Lambda ^2}}}{\Lambda^2}} \right)
\label{VCW}
\eea
where:
\bea
\Omega = \frac{\lambda }{{4!}}\Phi _c^4 - \frac{1}{2}B{\kern 1pt} \Phi _c^2 - \frac{\lambda }{{4!}}C{\kern 1pt} \Phi _c^4
\eea
Note, at the intermediate stage the UV divergences are 
regulated by introducing a cut-off, $\Lambda^2$, when performing the $k^2$ integration.
Thus, in the $\Lambda\rightarrow \infty$ limit, we have the CW  result:
\be
W_{eff} = \Omega + \frac{{\lambda {\Lambda ^2}}}{{64{\pi ^2}}}\Phi _c^2 
+ \frac{{{\lambda ^2}\Phi _c^4}}{{256{\pi ^2}}}\left( {\ln \frac{{\lambda \Phi _c^2}}{{2{\Lambda ^2}}} 
- \frac{1}{2}} \right)
\label{CW0}
\ee
Following CW, the renormalisation conditions are: 
\be
{\left. {\frac{{{d^2}W_{eff}}}{{d\Phi _c^2}}} \right|_{\Phi_c=0}} = 0,\;\;\;{\left. {\frac{{{d^4}W_{eff}}}{{d\Phi _c^4}}} \right|_{\Phi_c=M}} = \lambda ,\;\;\; \left. Z\right|_{{\Phi_c=M}} = 1
\ee
Here CW renormalise at an ``external" mass scale, $M$, 
to avoid  the IR singularity. 
Implementing these conditions\footnote{There is no wave-function renormalisation at 1-loop order} 
determines the counter terms and gives the final CW result:
\be
W = \frac{\lambda }{{4!}}\Phi_c^4 + \frac{{{\lambda ^2}\Phi _c^4}}{{256{\pi ^2}}}\left( {\ln \frac{{ \Phi _c^2}}{{{M^2}}} - \frac{{25}}{6}} \right)
\label{CWpot0}
\ee
In terms of the original fields $\Phi=f \Theta$, $\Theta\approx{\theta\over\sqrt{1-\alpha_1}}$
and $\theta\approx \hat\phi_1/\hat\phi_2$,
the potential is given by:
\be
W \approx \frac{\lambda }{{4!}} \hat\phi_1^4 + 
{\lambda ^2 \hat\phi^4_1\over 256\pi^2}\left( {\ln \left( {C\hat{\phi}^2_{1c}\over \hat{\phi}^2_{2c}}\right)- \frac{{25}}{6}} \right)
\label{CWpot}
\ee
where $C={f^2\over M^2}{1\over 1-\alpha_2}$ is a 
constant invariant under scale changes. 
This is the Weyl invariant CW potential written in
terms of the variables $(\hat\phi_1,\hat\phi_2)$ which are constrained by 
eq.(\ref{constrained}).  In addition there is a dilaton, $\sigma$, with
an isolated kinetic term. By performing a Weyl transformation
that is the inverse of eq.(\ref{W11}), we can relax
the constraint eq.(\ref{constrained}) and obtain, 
\be
W \approx \frac{\lambda }{{4!}} \phi_1^4 + 
{\lambda ^2 \phi^4_1\over 256\pi^2}\left( {\ln \left( {C{\phi}^2_{1c}\over {\phi}^2_{2c}}\right)- \frac{{25}}{6}} \right)
\label{relaxed}
\ee
which is Weyl invariant, and the the fields $(\phi_1,\phi_2)
=\exp(-\sigma/f)(\hat\phi_1,\hat\phi_2)$ are independent
variables.

The reason Weyl invariance has been preserved is because 
the inertial spontaneous symmetry breaking has introduced 
the mass scale, $f$, that compensates for the appearance 
of the renormalisation scale $M$ under the log, leaving the logarithmic terms invariant. 
Note that the usual renormalisation group equations still 
apply as a change in the renormalisation scale $M$ (a change in $C$ in eq.(\ref{CWpot})) 
is compensated by a change in the couplings and wave function factors in the usual way.

\subsubsection{Scale invariant regularisation}

The standard regularisation described above clearly preserves Weyl invariance even away 
from the small 
${\phi_1 \over \phi_2}$ limit because, on dimensional grounds,  
the spontaneous scale breaking factor, $f$, always compensates 
for the renormalisation scale factor to give an overall constant under 
the log, together with a   function of the scale invariant field $\Theta=f\Phi$. 

Expanding eq.(\ref{2scalarS}) beyond leading order leads to higher order terms in $\theta$ but 
these non-renormalisable terms are small. The reason is that Planck scale 
is predominantly due to the VEV of $\phi_2$ whereas the VEV of $\phi_1$, which 
models the SM Higgs,  is at the electroweak scale so that the non-renormalisable 
terms are Planck suppressed. In order to generate the hierarchy in the VEV's at 
the IR fixed point it is necessary that  the only large coupling is $\lambda$ while 
the other couplings associated with the other scale invariant quartic interactions are hierarchically 
small and can be neglected when calculating the radiative corrections. 

Of course there will 
be further terms when the gravitational interactions are included. 
Gravitational corrections require the addition of the Weyl tensor, $W^2$, and
$R^2$ terms, which are induced by matter loops and have logarithmically running
coefficients.  An analysis of the full renormalization group
equations appears in \cite{Strumia}.  While the Weyl tensor term is locally invariant, the $R^2$ term
is only globally invariant. Hence we expect to maintain a
conserved current, $K_\mu'$, however
the current will be modified by the addition of a new term,
$K'_\mu = K_\mu + c'\partial_\mu R/f_0^2$
in the notation of \cite{Strumia}.  We expect that this is a small correction
to the above scenario of a fixed ellipse, but may have some phenomenological implications
that will be pursued elsewhere. 

Another potentially challenging consequence of the gravitational corrections
is that the $\lambda_i$ become locked to the $\alpha_i$  by the renormalization group.
This may necessitate some large fine-tunings to maintain a small cosmological
constant and/or flat potentials.  We feel that this requires a more sophisticated fundamental analysis
since the RG equations computed in flat geometries amount ot a ``gauge choice'' for the Weyl symmetry
and do not admit analysis of the Weyl transformation. 

Finally, it is possible to maintain the local Weyl symmetry without choosing special
values of the $\alpha_i$, but rather by introducing the Weyl vector potential. 
When this is done, the dilaton is ``eaten'' to become the longitudinal part of a massive
Weyl vector potential.
The relationship of this to gravitational corrections  and our general framework is unexplored.

\subsubsection{Scale invariant dimensional regularisation}
Of course regularisation should not depend on the method used to control 
the intermediate divergences. Up to now we have used a momentum space 
cut-off but it is straightforward to use dimensional regularisation. In this case 
one first continues the theory to d-dimensions and introduces an external mass scale, 
$\mu$, to relate the 4-D dimensionless couplings to the dimension-full ones in d-dimensions. 
For the 2-scalar theory discussed above, dimensional regularisation leads straightforwardly 
to the form of eq.(\ref{CWpot}) with $M$ replaced by $\mu$. In this case the quartic 
and quadratic terms are automatically absent. The dependence on the mass parameter, 
$\mu$, needed to continue away from four dimensions, will always appear in the scale invariant 
ratio $\mu/f$  giving eq.(\ref{CWpot}) as before.

\subsubsection{Relation to previous regularisation proposals}
Scale invariant dimensional regularisation that differs from the one just described has 
been considered by several authors 
\cite{Englert:1976ep}\cite{S3}.
The method generally adopted to maintain scale invariance in radiative order replaces $\mu$ 
by a function of the scalar fields, $\mu\rightarrow \mu(\phi_i)$, with the appropriate scaling behaviour.
In this case  the $d$-dimensional tree level potential $\tilde V$ has the form
\be
\widetilde V\left( {\phi ,\chi } \right) \equiv \mu{\left( {\phi ,\chi } \right)^{4 - d}}V\left( {\phi ,\chi } \right).
\label{VT}
\ee
 As a result the tree level potential introduced in eq.(\ref{VT}) has {\it additional} interactions of the form
\be
\tilde W(\phi,\chi)-W(\phi,\chi)=(4-d)\;W(\phi,\chi)\;\ln\;\mu(\phi,\chi)+O(4-d)^2.
\label{Vdim}
\ee 

Although these interactions vanish in 4 dimensions, they give a finite correction to $W_{eff}$ at 
1-loop order because the underlying divergence in 4-dimensions  cancels the $4-d$ factor in the 
additional term in eq.(\ref{Vdim}). Thus, due to the additional interaction terms in eq.(\ref{Vdim}) 
that depend on the choice of $\mu(\phi,\chi)$, the scale invariant d-dimensional theory is {\it not} 
the same as that defined purely in 4-dimensions. As a result the final regulated theory in 4-dimensional 
has additional terms that depend on the precise choice of the regulator $\mu(\phi,\chi)$. For the 2-scalar
case with potential given by eq.(\ref{VT}) and the choice $\mu(\phi,\chi)=\chi$ the additional term at one-loop
is of the form $\phi^6/\chi^2$. While this is still scale invariant it means the resulting 4-dimensional potential
is different from that obtained by the regularisation procedure discussed above. The origin of this discrepancy is that the requirement that scale invariance be preserved in d-dimensions rather than regularisation ambiguity requires such additional terms and defines a different theory. 

In summary, we have shown that the standard regularisation procedure preserves scale invariance.
It does not involve the introduction of an arbitrary regularisation function and, although it involves 
non-renormalisable interactions, these are well defined. Of course it is possible to add additional 
non-polynomial terms to the theory while preserving scale invariance but we see no reason to do so.

\section{Summary and conclusions}\label{ssummary}

We have discussed how inflation  and Planck  scale generation  
can emerge  from a dynamics  associated  with  global Weyl  
symmetry and its current,  $K_{\mu}$.    
In  the   pre-inflationary universe,  the  Weyl  
current  density,   $K_0$ ,  is driven  to  zero by general  expansion.   
However,  $K_{\mu}$  has a kernel structure, i.e.,  $K_{\mu}=\partial_{\mu}K$ 
and, as $K_0\rightarrow 0$, the kernel  evolves as $K\rightarrow\bar K$, constant. 
This  resulting  constant $\bar K$, that does not depend on the scalar potential, 
is the  order  parameter of the  Weyl  symmetry breaking; 
indeed,  $\bar K$  directly  defines the Planck mass. 

In $N$-multi-scalar-field theories $K$ has the general
form $K=-\frac{1}{2}(F(\{\phi_j\})- \sum_{i=1}^N\phi^2_i $
for nonminimal coupling $-(1/12)F(\{\phi_j\})R$. The fields become
constrained to the manifold $K\rightarrow \bar{K}(\{\phi_j\})$.
In detail we have studied
$F(\{\phi_j\})=\frac{1}{2}\sum_{i=1}^N\alpha_i\phi^2_i $.
This defines an ellipsoidal 
constraint on the scalar field VEVs.  An inflationary slow-roll period 
is then  associated with  the  field VEVs  migrating along  the  ellipse.
Up to this point the fate of scale symmetry is entirely controlled
by the inertial symmetry breaking, $K\rightarrow \bar{K}(\{\phi_j\})$.
 A potential ultimately 
sculpts the ensuing slow roll on the manifold to the IR, and defines  the ultimate
vacuum (together with any quantum effects that may distort the $K$ ellipse \cite{FHR}) 
This fixes the relative value of the scalar field VEVs through quartic terms only.
There is a harmless massless dilaton associated with the dynamical symmetry breaking
which represents dilations of the ellipsoid.  We emphasize that with more
general choices of $F(\{\phi_j\})$, the constraint manifold can become 
a more general manifold in the field space, and it would be of interest to explore the
possibilities in this case.

Any  Weyl  symmetry breaking  effect at the quantum level is intolerable  and 
will show up as a nonzero divergence in the $K_{\mu}$ current.  We showed how, due 
to the decoupling of the dilaton,  these quantum effects actually preserve the Weyl 
symmetry using the normal momentum space cut-off or dimensional regularisation schemes. 
The potential scale dependence introduced by the  ``external" mass scale needed to regulate 
the logarithmic divergences is cancelled by the scale invariant order parameter responsible 
for spontaneous breaking of the Weyl symmetry.  It would be of interest to
study the local Wetyl invariant theories that involve the Weyl photon, as in
Section II.C, in great detail. This provides an example of an inertial Higgs mechanism,
and the dilaton is eaten and completely removed from the low energy spectrum.

A strong motivation for considering such Weyl invariant theories is to provide a solution to 
the hierarchy problem of the Standard Model. In the absence of gravity or very massive 
states associated with the Landau pole of the Standard Model or of an extension of the 
Standard Model such as Grand or string unification, the Standard Model is natural in the sense 
that the quadratic divergence found in radiative corrections to the Higgs mass is unphysical and 
is cancelled by the mass counter term. Requiring scale invariance ensures that the Higgs is massless  
but, of course, some mechanism to spontaneously break the scale symmetry is needed. 

If gravity 
is included via the Weyl invariant extension discussed here, then the Standard Model {\it plus} gravity 
is natural in the sense just discussed. Of course it is still necessary that there be no 
massive states strongly coupled to the Higgs with masses much larger than the electroweak scale.  
Moreover the scale symmetry is now automatically spontaneously broken by the inertial mechanism. 
To obtain the hierarchy between the Planck scale and the electroweak breaking scale it is necessary 
to have hierarchically large ratios of the dimensionless couplings of the scalar potential. In the 
absence of gravitational radiative corrections, these ratios are only multiplicatively changed by radiative 
corrections and thus are natural. This may be seen from the underlying shift symmetry of the Weyl 
invariant Higgs potential. 

This shift symmetry is broken by the Higgs coupling to the Ricci scalar. To determine 
whether the hierarchy is ultimately preserved requires a calculation of the gravitational radiative 
corrections which is beyond the scope of the present paper. In a Weyl invariant variation of the 
Standard Model with no gravity, no grand unification and no Landau poles in the far 
UV the Higgs mass 
is technically natural with no  hierarchy problem!

\vskip 0.5 in
\noindent
 {\bf Acknowledgements}
\vspace{0.1in}

We thank W. Bardeen,  D. Ghilencea, A.~Salvio, and A.~Strumia for discussions. 
PGF acknowledges support from
STFC, the Beecroft Trust and the ERC. Part of this work was done at 
Fermilab, operated by Fermi Research Alliance, 
LLC under Contract No. DE-AC02-07CH11359 with the United States 
Department of Energy. 

\end{document}

\bibitem{CCJ} 
  C.~G.~Callan, Jr., S.~R.~Coleman and R.~Jackiw,
  Annals Phys.\  {\bf 59}, 42 (1970).
  doi:10.1016/0003-4916(70)90394-5

\end{document}